\documentclass[11pt,twocolumn]{article}

\usepackage{times}

\usepackage[labelfont=bf]{caption}

\usepackage[top=0.5in,bottom=0.75in,left=0.5in,right=0.5in]{geometry}

\usepackage{url}

\usepackage{subfig}

\usepackage{graphicx}

\usepackage{amsmath}

\usepackage[numbers]{natbib}
\usepackage[bookmarksopen,bookmarksnumbered]{hyperref}

\usepackage[compact]{titlesec}

\usepackage{helvet}
\usepackage{sectsty}
\allsectionsfont{\sffamily}

\begin{document}


\title{The Stories We Tell About Data:\\ Media Types for Data-Driven Storytelling}

\author{Zhenpeng Zhao and Niklas Elmqvist\footnote{Corresponding author: \url{elm@umd.edu}.}\\ \scriptsize University of Maryland, College Park}

\date{January 2022}

\maketitle

\section*{Abstract}

The emerging practice of data-driven storytelling is framing data using familiar narrative mechanisms such as slideshows, videos, and comics to make even highly complex phenomena understandable.
However, current data stories are still not utilizing the full potential of the storytelling domain.
One reason for this is that current data-driven storytelling practice does not leverage the full repertoire of media that can be used for storytelling, such as the spoken word, e-learning, and video games.
In this paper, we propose a taxonomy focused specifically on media types for the purpose of widening the purview of data-driven storytelling simply by putting more tools into the hands of designers.
Using our taxonomy as a generative tool, we also propose three novel storytelling mechanisms, including for live-streaming, gesture-driven oral presentations, and textual reports that dynamically incorporate visual representations.

\textbf{Keywords:} Data-driven storytelling; narrative visualization; taxonomy; design guidelines.
    
\begin{figure*}[tbh]
  \centering
  \subfloat[Annotated chart created in \textit{ChartAccent}.]{
    \resizebox{!}{4cm}{\includegraphics{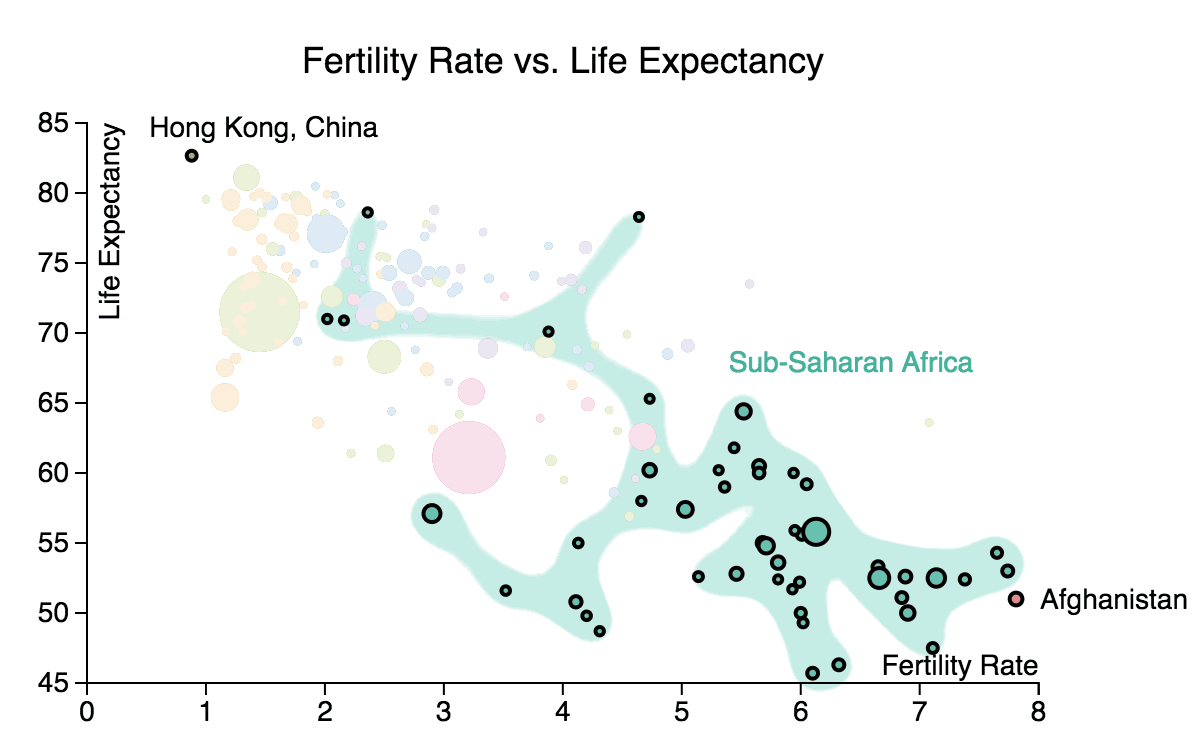}}
    \label{fig:chartaccent}
  }
  \subfloat[Data-driven story of a session in the video game \textit{Civilization 3} (2001).]{
    \resizebox{!}{4cm}{\includegraphics{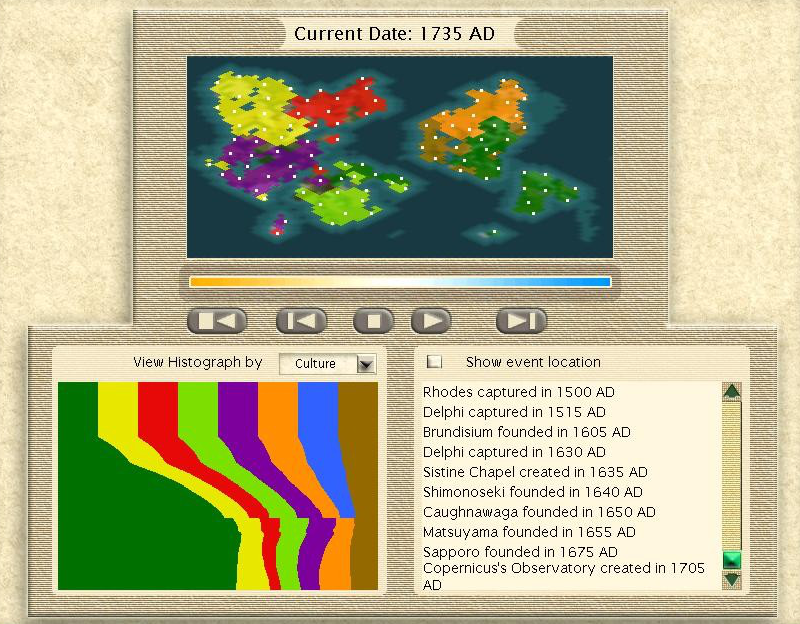}}
    \label{fig:civ3-replay}
  }
  \subfloat[Will Ferrell annotating everyday activities in \textit{Stranger Than Fiction} (2006).]{
    \resizebox{!}{4cm}{\includegraphics{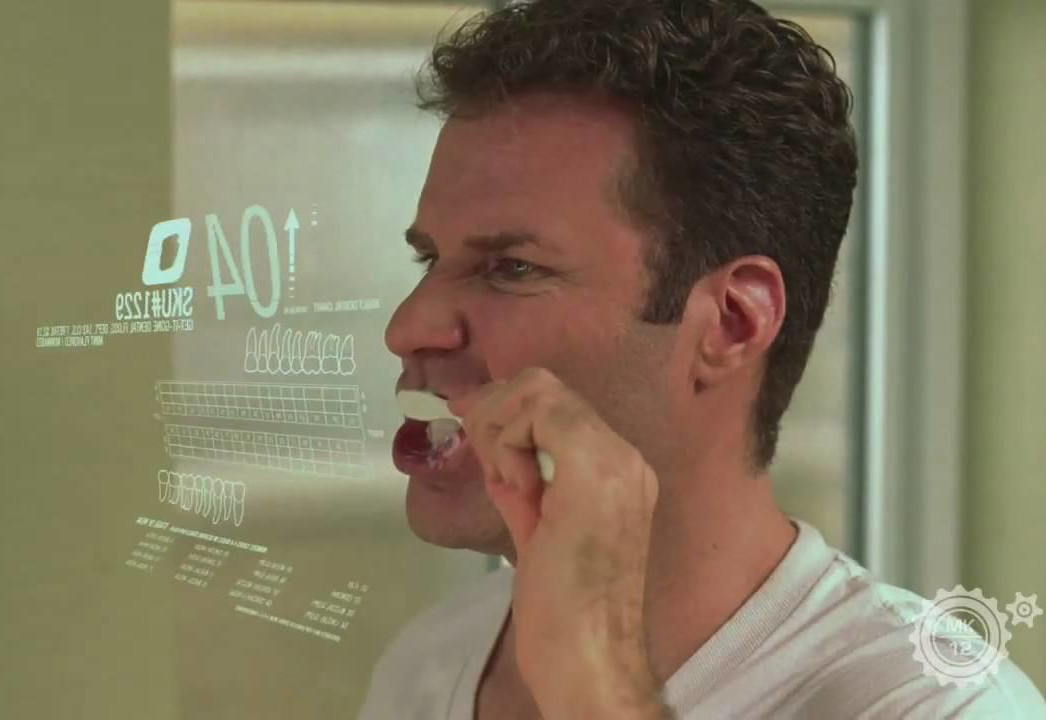}}
    \label{fig:stranger}
  }
  \caption{Many possible media types are available today for data-driven storytelling.}
  \label{fig:teaser}
\end{figure*}

\section{Introduction}

From the hunter returning from his latest foray to tell tall tales of stalking his prey, to the shaman spinning a yarn about the origins of the gods, the stars, and the moon, \textit{storytelling} is the oldest form of human communication, record-keeping, and entertainment.~\cite{Leitch1986, Schank1995, Vansina1985}
\textit{Stories}---sequences of events involving characters and places---are particularly well suited for this purpose because their chronological structure enables recall, entices listeners, and facilitates understanding.~\cite{Gottschall2012, Leitch1986}
For this reason, narration and storytelling retain important roles even in today's information society, where these properties are particularly important in helping people get to grips with an increasingly complex world. 
This has recently given rise to \textit{data-driven storytelling} where narrative techniques are utilized for telling stories about data~\cite{Segel2010}, often using visual media.~\cite{Eisner2008, Sless1981} 

Examples of such data-driven storytelling abound, and include data comics,~\cite{Bach2017, Zhao2019} infographics,~\cite{Harris1999} data videos,~\cite{Amini2015, Amini2017} data-driven slideshows,~\cite{Kosara2013} and even story sketching.~\cite{Lee2013}
The common denominator is that they are based on story arcs evolving over time to build an argument, explain a phenomenon, or report a finding.
They diverge by the form of \textit{media}, or channels of communication, they use: sequential art, animated graphics, presentation slides, and handwriting, respectively.

However, while visual representations are particularly useful for storytelling, there are many additional types of media---digital and otherwise---that can be co-opted for telling stories about data.
An existing framework proposed by Segel and Heer~\cite{Segel2010} outline seven \textit{genres} of narrative visualization, but conflates the storytelling mechanism with the media being used.
Furthermore, Segel and Heer also note that their sample is limited (58 items) and does not cover the full scope of possible media that can be used for storytelling,~\cite{Segel2010} such as video games, infotainment, and other e-learning tools.
Meanwhile, much research in data-driven storytelling has used these seven genres as a starting point, which suggests that the community may be limiting itself by needlessly adhering to a framework that was intended to be generative rather than prescriptive.

What about using the spoken word for data-driven storytelling, i.e., supporting speakers talking to an audience?
What about the written word, i.e., data-driven prose, such as for inclusion in a textual report?
And what about other forms of human communication, such as theatre, poetry, and even dance?
Much innovation remains in data-driven storytelling, but this requires going beyond existing frameworks.
Hence this work.

We present a new taxonomy focused on media for data-driven storytelling with the purpose of opening the field to a wider set of future possibilities.
Our work started with collecting evidence of data-driven storytelling using novel and diverse media, from the spoken word to interpretative dance and choreography. 
We then use this wealth of data to derive a taxonomy and classify all of these examples into a coherent framework.
We use the framework to postulate some potential future media types for data-driven storytelling.
Finally, by generalizing across storytelling practice for different media, we derive design guidelines for data-driven storytelling.
We conclude the paper with a summary and our plans for future work.

\section{Data-Driven Storytelling}

\textit{Storytelling} is the conveyance of a sequence of events, often involving characters and places---\textit{stories}---using speech, sound, and visuals,~\cite{Gottschall2012} and has a history spanning thousands of years.~\cite{Schank1995, Vansina1985}
A \textit{visual narrative} is a story told primarily using visual media, such as illustrations, photographs, animations, video, and---now---visualization.~\cite{Eisner2008, Sless1981}
In particular, visualization has a specific proclivity for communication by virtue of its graphical form, yielding the notion of \textit{communication-minded visualization}~\cite{Viegas2006} to support collaborative analysis.
Combining the idea of communication-driven visualization with storytelling yields the notion of \textit{data-driven storytelling}: narrative techniques for telling stories about data.~\cite{Segel2010}

\begin{figure}[htb]
	\centering
    \resizebox{\columnwidth}{!}{\includegraphics{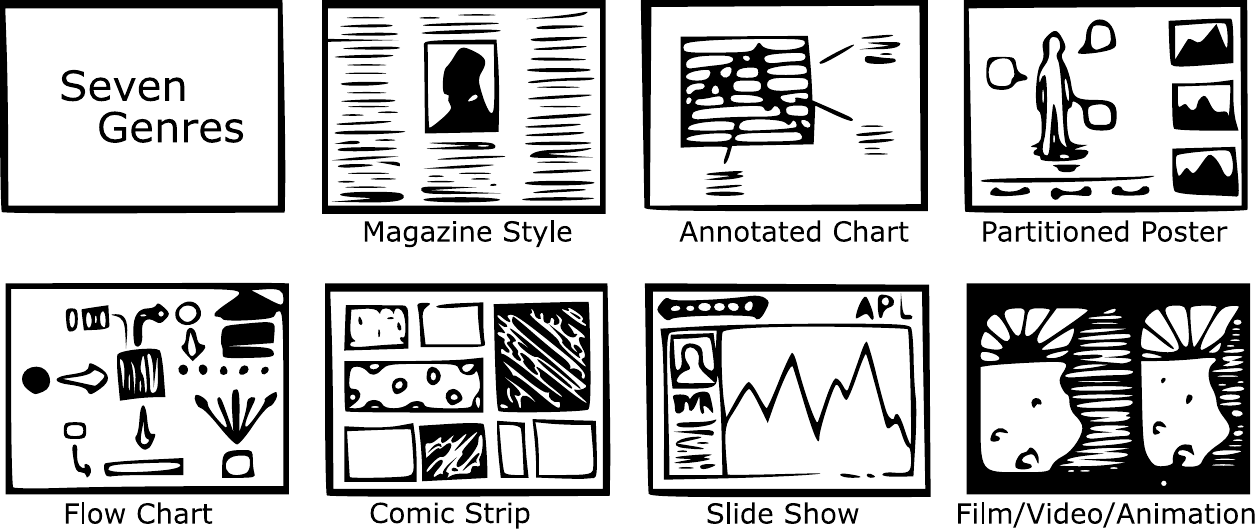}}
    \caption[]{Seven genres of narrative visualization by Segel and Heer.~\cite{Segel2010}}
    \label{fig:genres}
\end{figure}

\subsection{Existing Media Types}

Our focus in this paper is on communication mechanisms---or \textit{media}---used for conveying data-driven stories. 
In this paper, we define ``media'' as the channel or the tools used to store and deliver information. 
While data-driven storytelling is a nascent research topic in visual communication and visualization, there has so far been no focus directed to the specific media used.
Instead, existing efforts tend to revolve around the seven genres of narrative visualization proposed by Segel and Heer~\cite{Segel2010} (Figure~\ref{fig:genres}):

\begin{itemize}
    \item\textbf{Magazine Style:} A data-driven image integrated in a page of text, where the text refers to and explains the image.

    \item\textbf{Annotated Chart:} Chart adorned with descriptive text and labels for the purpose of explaining its contents.

    \item\textbf{Partitioned Poster:} A poster or dashboard consisting of multiple images, each with separate data.

    \item\textbf{Flow Chart:} Visually ordered sequence of images and annotations designed to tell a story.

    \item\textbf{Comic Strip:} Sequence of frames containing images and text organized in a comic-style strip layout.

    \item\textbf{Slide Show:} Deck of slides combining images and text to sequentially tell a story.

    \item\textbf{Film/Video/Animation:} Motion graphics that incorporate data-driven imagery and visualizations, often animated.
\end{itemize}

However, as readily admitted by Segel and Heer, their findings are limited to a sample of 58 examples.
They also do not claim that their genres are exhaustive, noting for example that their work did not include video games or e-learning tools. 
Furthermore, the above seven genres conflate the media used for storytelling with the format, method, and components employed. 
The seven genres have proven to be extremely powerful for categorizing research in this field; they have even played a prescriptive role, with Graph Comics~\cite{Bach2016} and Data Comics~\cite{Zhao2019} arising as examples of the Comic Strip genre, and Data Videos~\cite{Amini2015, Amini2017} drawing inspiration from the Film/Video/Animation genre.
However, there certainly is room for expanding the framework further.

\subsection{Visual Communication}

Visual forms of communication, including icons, illustrations, schematics, photographs, and full-motion video, have long been considered to be one of the most compelling and approachable storytelling formats.~\cite{Chevalier2016, Sless1981}
Presenting insights from data to the masses requires taking the visualization literacy~\cite{Boy2014} of everyday users into account.
Thus, the notion of ``casual visualization''~\cite{Pousman2007} is important.

Complex skills are today often learned through step-by-step instructions on the internet.~\cite{Grossman2010}
Several projects have endeavored to increase the richness of such video tutorials: ToolClips~\cite{Grossman2010} provide contextual access to video assistance, Pause-and-Play~\cite{Pongnumkul2011} links tutorials to the real application, MixT~\cite{Chi2012} automatically generates video from demonstrations, and a recent approach makes software tutorial videos interactive.~\cite{Nguyen2015}
  
Beyond traditional providers (HBO, MSNBC, ABC, etc) as well as the new breed (Amazon Prime, Hulu, and Netflix), open video sharing sites such as YouTube have democratized access to visual storytelling.
With such openness comes entirely new storytelling formats that were previously infeasible, such as vlogs (video blogs),~\cite{Griffith2010} reaction videos,~\cite{Anderson2011} and video commentaries, including the ever-popular Let's Play videos.~\cite{Wadeson2013}
An additional recent development in internet video is the increased focus on live, so-called \textit{streaming}, content, where the media is received and presented to the consumer at the same time it is delivered by the provider.
This development was particularly driven by live gameplay content, where prominent gamers---called \textit{streamers}---broadcast their game screen on services such as Twitch.tv.
While streaming is also used for heavily produced events such as eSports tournaments, where professional gamers compete over thousands of dollars in prize money in games such as \textit{Overwatch}, \textit{League of Legends}, and \textit{Dota 2}, the vast majority of live streams on Twitch and elsewhere are created using specialized \textit{broadcasting software} such as OBS, XSplit, GameShow, etc.

Amini et al.~\cite{Amini2015} recently identified \textit{data videos} as motion graphics combining both sound and visuals to tell a data story.
Pointing to prominent examples from the New York Times and the Guardian, their work encourages professional storytellers to use visuals to craft their narratives.
Their follow-up Data Clips~\cite{Amini2017} work is an authoring tool for creating data-driven clips incorporating visualizations that can be assembled into longer data videos.

\subsection{Visualization for Communication}

Data visualization is the use of interactive graphical representations of data to aid cognition.~\cite{Card1999}
There are arguably two main uses of data visualization~\cite{Viegas2006}:
\begin{itemize}
    \item\textbf{Exploration:} Visualization for {exploratory data analysis}~\cite{Tukey1977} to gain insights and generate hypotheses; and
    
    \item\textbf{Explanation:} Visualization for informing external stakeholders about data using visual means.
\end{itemize}

Data-driven storytelling is a natural extension of visualization for explanation (the latter).
The production, presentation, and dissemination of results is a grand challenge of visualization and visual analytics.~\cite{Thomas2005}
Gershon and Page first proposed using storytelling for visualization~\cite{Gershon2001}, and their work has since been followed up by workshops,~\cite{Diakopoulos2011, DiMicco2010} surveys,~\cite{Hullman2011, Segel2010} and even commercial tools.~\cite{Kosara2013}
Vi{\'e}gas and Wattenberg remark upon the proclivity of visualization for communication by virtue of its graphical form, and encourage focusing on so-called \textit{communication-minded visualization}~\cite{Viegas2006} where communication enables collaborative analysis.

To reach its full potential, communication capabilities should be integrated into the visualization tools themselves~\cite{Thomas2005}; for example, Tableau incorporates story points,~\cite{Kosara2013} and most commercial tools support exporting interactive dashboards to the web.

\section{Evidence of Data-Driven Storytelling Media}

To illustrate the prevalence and variety of different data-driven narratives in the world, we here enumerate and discuss a set of representative and innovative such examples.
The purpose is to provide a basis for a taxonomy that can be used to classify the storytelling media used for the data-driven narrative.
For each example, we use an informal classification scheme to describe the media in more detail. 
This scheme will then feed into our taxonomy in the following section.

We use the following criteria for our selection:
\begin{itemize}
    \item Has a \textit{story} format, i.e., a progression (or arc) over the course of the artifact;
    \item Leverages \textit{data} to drive its narrative;
    \item Uses an approach that is \textit{novel} over Segel and Heer's seven genres~\cite{Segel2010}; and
    \item Inhabits a \textit{unique} position w.r.t\ other examples. 
\end{itemize}

\begin{figure}[htb]
	\centering
    \resizebox{\columnwidth}{!}{\framebox{\includegraphics{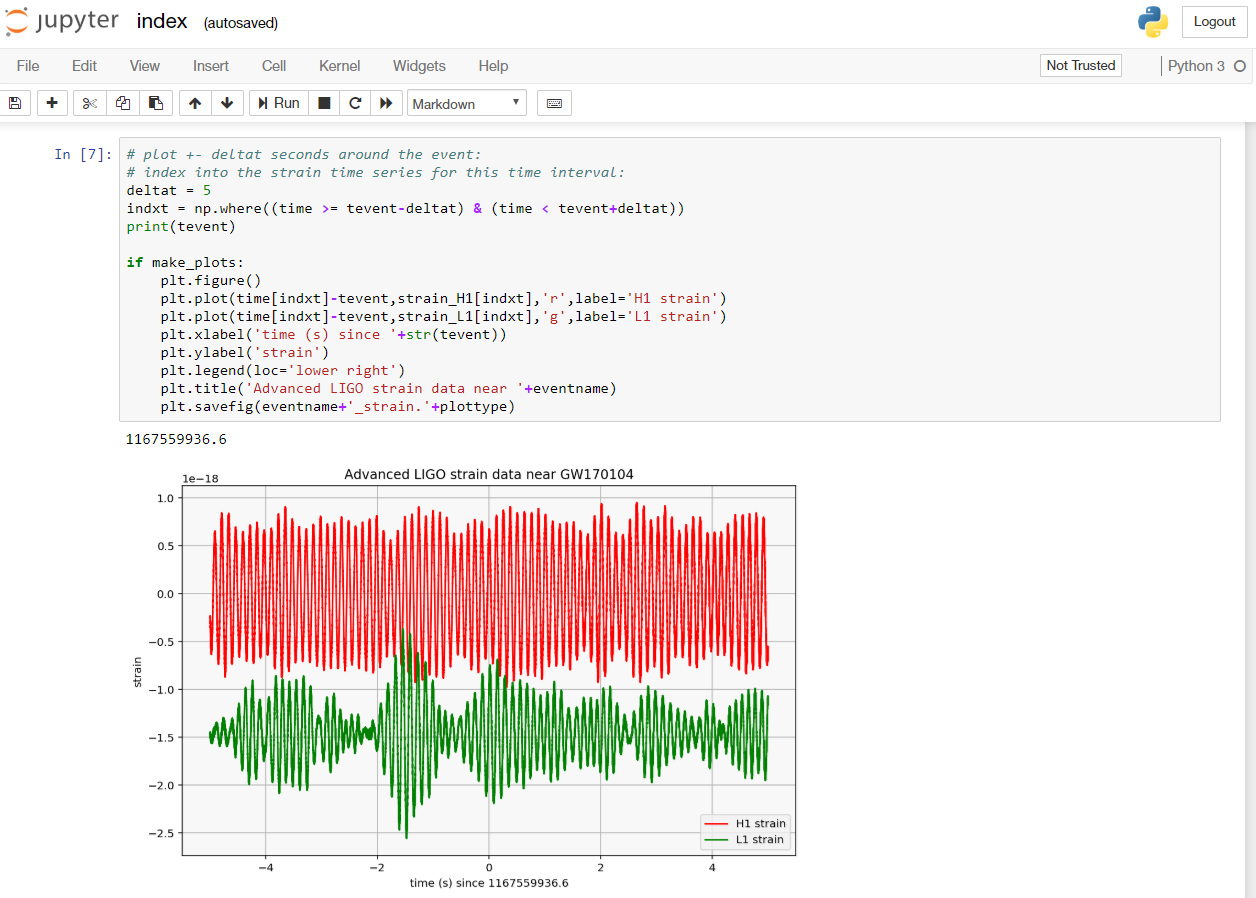}}}
    \caption[]{Interactive signal processing of the LIGO-Virgo collaboratory data for detecting gravitational waves in a Jupyter notebook~\cite{kluyver2016jupyter}.}
    \label{fig:jupyter}
\end{figure}

\begin{figure}[htb]
	\centering
    \resizebox{\columnwidth}{!}{\framebox{\includegraphics{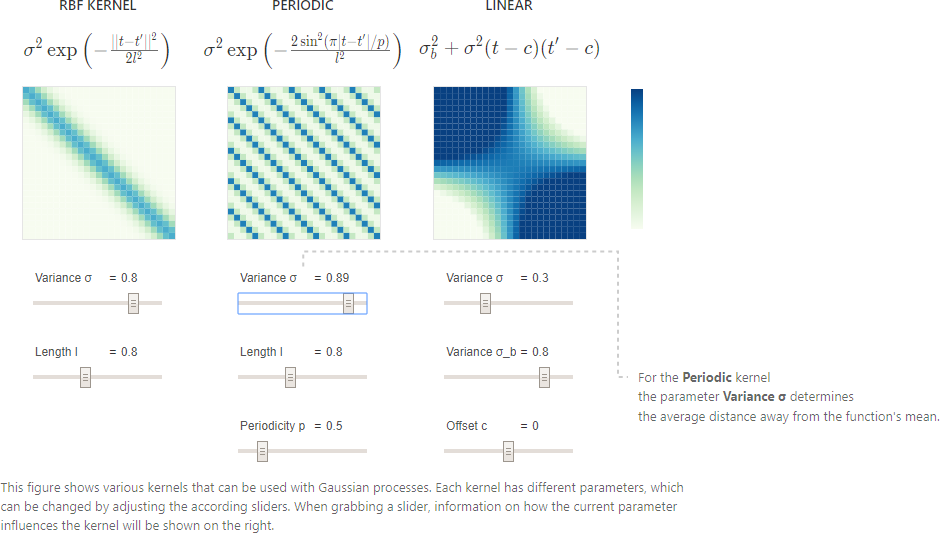}}}
    \caption[]{Interactive article on Gaussian processes~\cite{Gortler2019a} from \textit{Distill}, an online machine learning journal.
    Manipulating widgets will modify text and figures accordingly.
    }
    \label{fig:distill}
\end{figure}

\begin{figure}[htb]
	\centering
    \resizebox{\columnwidth}{!}{\framebox{\includegraphics{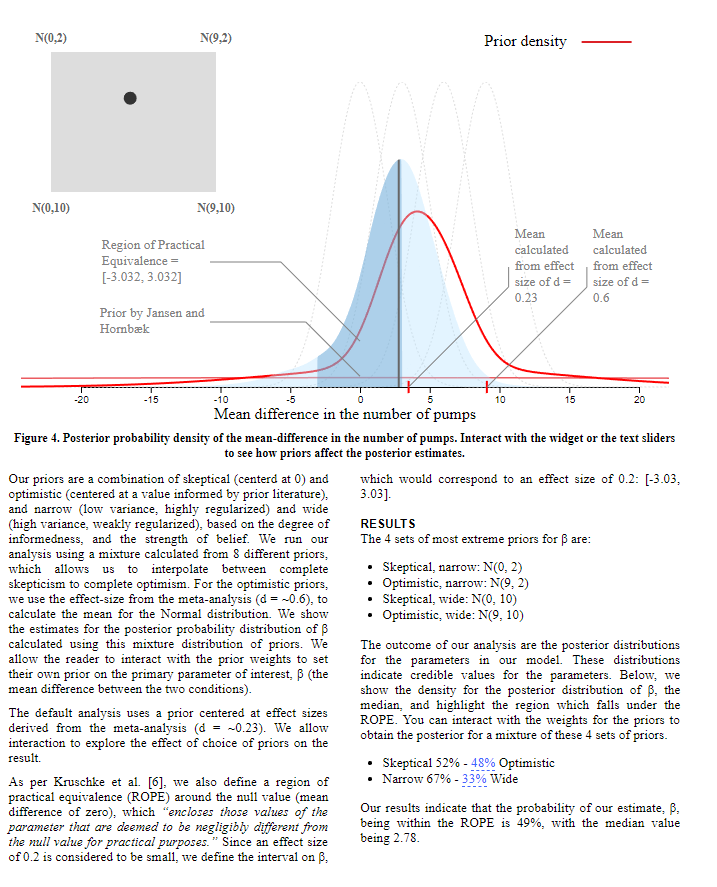}}}
    \caption[]{An explorable report for multiverse analysis, as proposed by Dragicevic et al.~\cite{Dragicevic2019}
    Clicking underlined parameters in the text toggles between different animations and changes the visual representations.
    }
    \label{fig:multiverse}
\end{figure}

\begin{figure}[htb]
	\centering
    \resizebox{\columnwidth}{!}{\framebox{\includegraphics{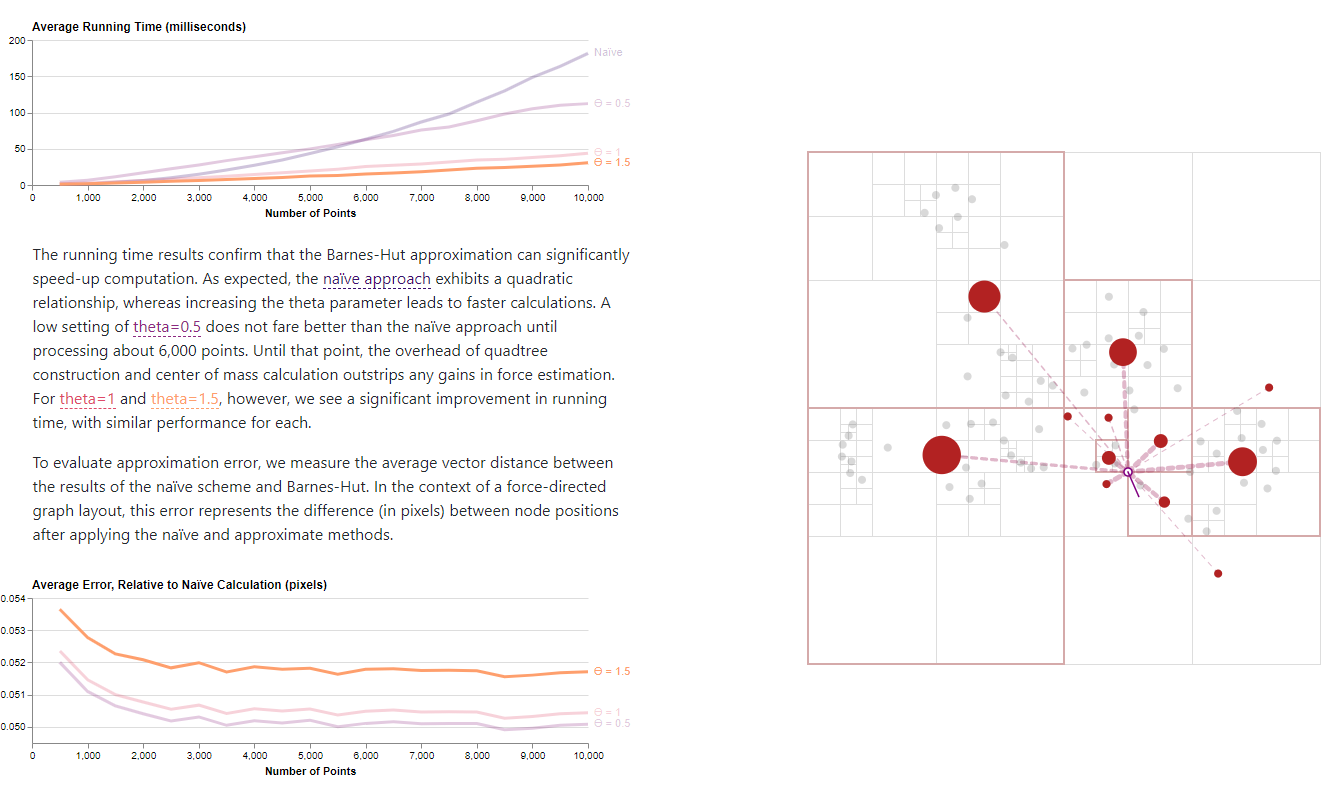}}}
    \caption[]{Jeffrey Heer's interactive article (written in Idyll~\cite{Conlen2018}) explaining the Barnes-Hut approximation for graph layout.
    As is typical for this type of artifact, the article contains multiple widgets that allows the reader interact with the displayed model.
    }
    \label{fig:idyll}
\end{figure}

\subsection{Exhibit 1: Interactive Articles}

Online journalism is changing to the point where so called \textit{interactive articles} are becoming a new, popular, and dominant form of visual communication.~\cite{Conlen2018}
These data-driven narratives---or \textit{explorable explanations}~\cite{Victor2011}---combine traditional journalistic storytelling with interactive components that allow a potentially large audience to engage with them.
Major newspapers such as the \textit{New York Times}, \textit{The Guardian}, and the \textit{Washington Post} have long made such interactive articles a popular part of their online presence; in fact, the New York Times graphics department was long a key driver for the D3~\cite{Bostock2011} toolkit (while Michael Bostock was a member until 2015).

While already tremendously successful in practice---ten of the New York Times' forty most popular articles in 2014 were interactive articles from its \textit{Upshot} department~\cite{Conlen2018}---the concept of interactive articles is also beginning to take hold in the academic realm.
Already in 2011, publishing giant Elsevier launched the Executable Paper Grand Challenge\footnote{\url{https://www.journals.elsevier.com/international-journal-of-human-computer-studies/news/introducing-executable-papers}} to find ways to improve reproducibility of research by including executable code inside an interactive article. 
The winner of this challenge was the Collage authoring environment,~\cite{NowakowskiCHKKBMDM11} and the Elsevier journal \textit{Computers \& Graphics} even published a special issue testing the environment,~\cite{SpagnuoloV13} but unfortunately, the movement seems to have died down within Elsevier.

Instead of starting from the article and going to interactivity, what about starting from code and going towards an article?
As embodiments of Knuth's \textit{literate programming}~\cite{knuth1984literate} paradigm, where traditional source code is embedded in descriptive natural language, \textit{computational notebooks}~\cite{colaboratory, observable, kluyver2016jupyter, radle2017codestrates, tabard2008individual, Wickham2017}---which combine executable code, its output, and descriptive text and other media in a single document---is based on this premise.
As a result, computational notebooks have become a \textit{lingua franca} for presenting findings, supporting material, and breakthroughs in many STEM disciplines~\cite{Rule2019}; for example, when gravitational waves were first observed by the LIGO-Virgo collaboration in 2015, the announcement in February 2016 was accompanied by a Jupyter notebook complete with all the collected data.\footnote{\url{https://www.gw-openscience.org/}}
However, computational notebooks are generally rather technical and expose interactive mostly through source code rather than widgets and controls. 

\textit{Distill},\footnote{\url{https://distill.pub/}} on the other hand, is an online academic journal for machine learning research that is based on interactive, explorable articles.
Figure~\ref{fig:distill} shows an example of a Distill article on understanding Gaussian processes where smooth controls allows the reader to change parameters, causing the visual representations to update accordingly. 
Similarly, the Dragicevic et al.~\cite{Dragicevic2019} propose \textit{explorable reports} that are designed specifically for multiverse analysis, where clickable controls in a paper allows for varying the reporting for many different statistical analyses in order to show the fragility or robustness of the findings.
Figure~\ref{fig:multiverse} shows an example of an example ``mini-paper'' implemented using this method; clicking on the blue underlined text will cycle between matching results and images.

Most recently and inspired by these efforts, Conlen and Heer~\cite{Conlen2018} proposed \textit{Idyll}, a domain-specific language for authoring interactive articles that is specifically targeted at journalists and designers. 
Using a rich standard library of components, Idyll allows even non-technical users to quickly create compelling and highly interactive content.
Figure~\ref{fig:idyll} shows an example interactive article created in Idyll on the Barnes-Hut approximation for graph layouts.\footnote{\url{https://jheer.github.io/barnes-hut/}}
In fact, Matthew Conlen, the creator of Idyll, has also founded \textit{Parametric Press}\footnote{\url{https://parametric.press/}}, an online magazine for interactive articles built in Idyll.

\textit{Informal classification:} 
Key in all of the interactive articles reviewed above is that they are \textit{interactive}; they allow---even \textit{invite}---interaction by the reader.
In some cases, this is done through low-level and technical means, such as changing source code in a Jupyter notebook, but in more ``polished'' cases intended for a more general audience, such as Distill and Idyll, the interaction is performed through standard widgets.

The other significant characteristic of interactive articles is inherent in the actual medium itself: they are based on a typical document format where the main context is textual in nature.
As with previous examples of new media being introduced into society, it is clear that data-driven stories will mostly augment rather than replace existing media types. 
In other words, a fruitful approach to invent future storytelling media for data is to study existing ones.

\begin{figure}[htb]
	\centering
    \resizebox{\columnwidth}{!}{\includegraphics{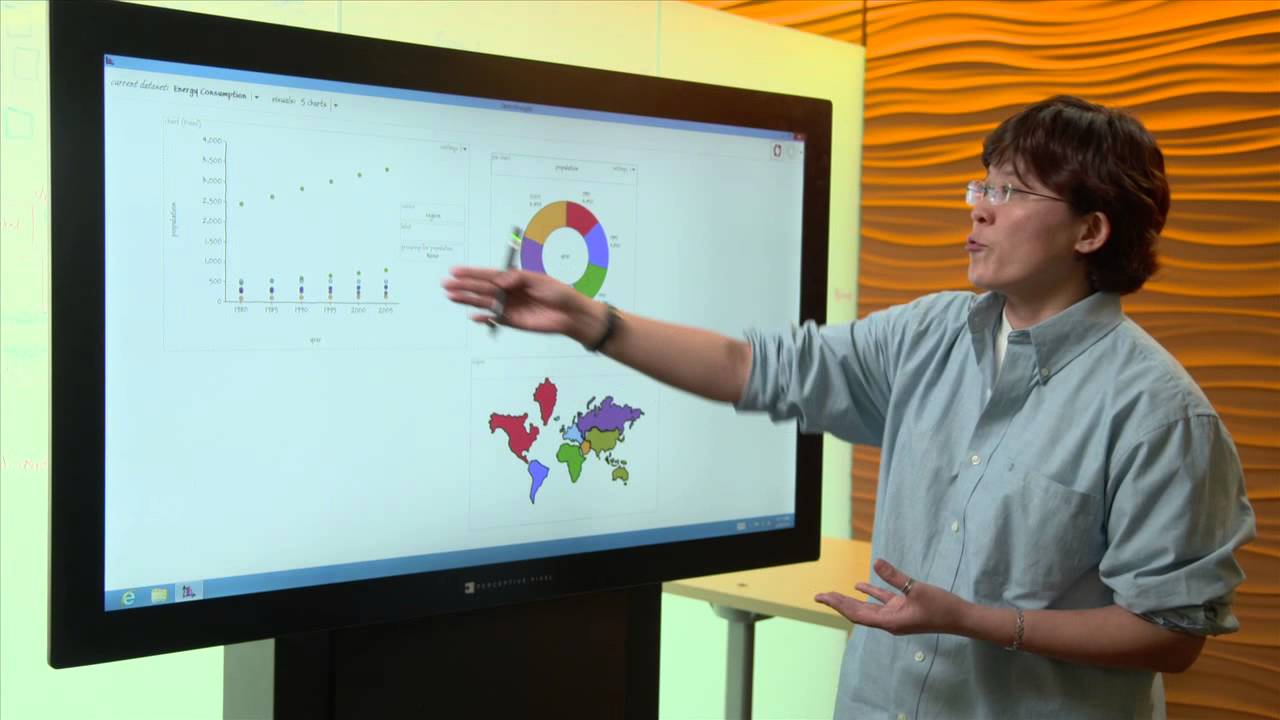}}
	\caption[]{Using SketchStory~\cite{Lee2013} to give an interactive and visual presentation on data.}
    \label{fig:sketch-story}
\end{figure}

\begin{figure}[htb]
	\centering
    \resizebox{\columnwidth}{!}{\includegraphics{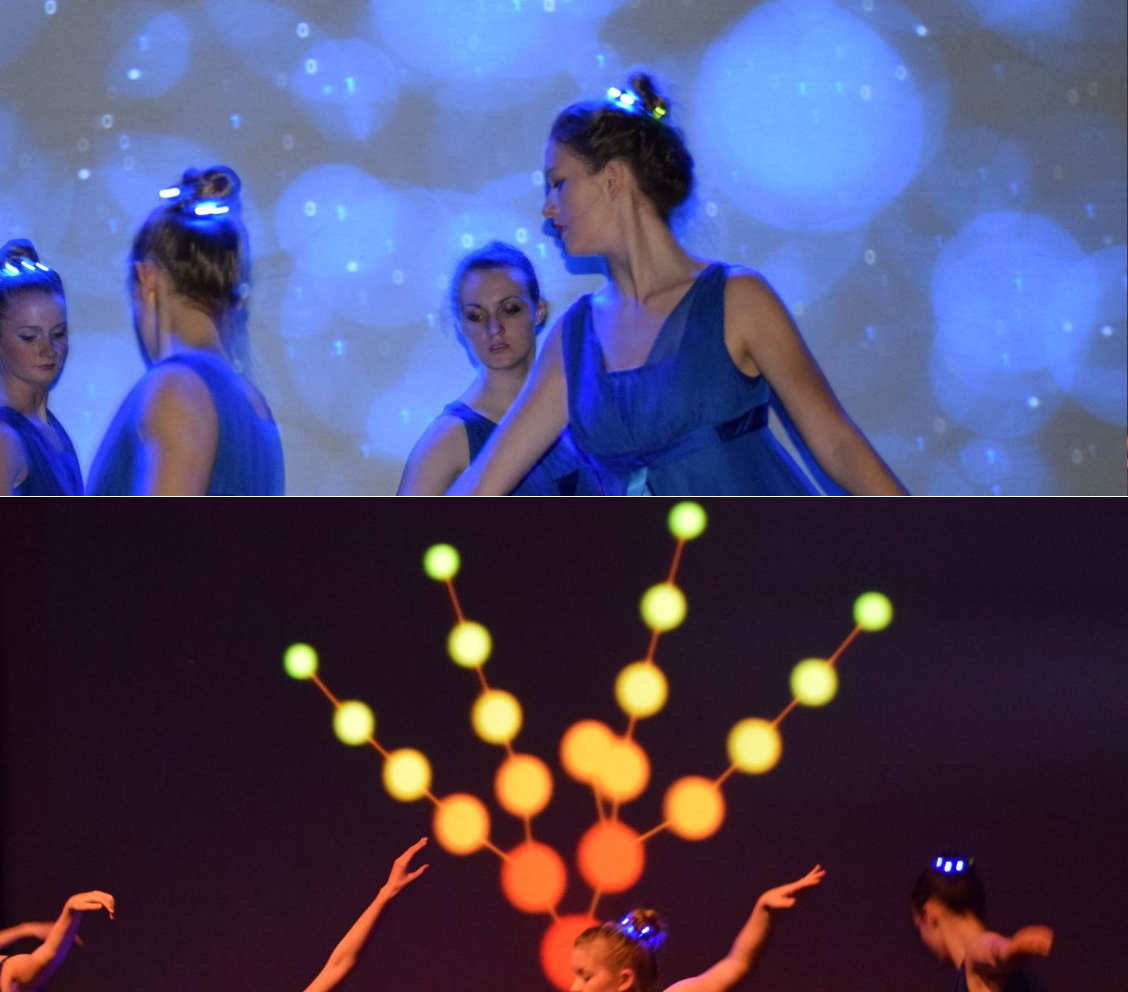}}
	\caption{Explaining computer science concepts using ballet.}
    \label{fig:arrastre}
\end{figure}

\begin{figure}[htb]
	\centering
    \resizebox{\columnwidth}{!}{\includegraphics{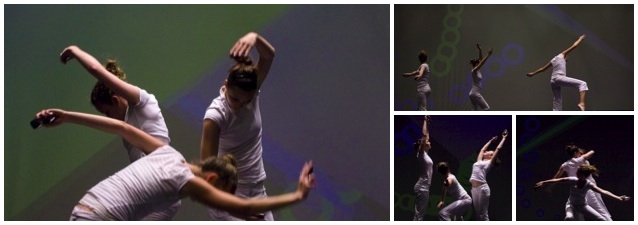}}
	\caption[]{Visualizing movement data in dance.draw~\cite{LatulipeH08}.}
    \label{fig:dance.draw}
\end{figure}

\begin{figure}[htb]
	\centering
    \resizebox{\columnwidth}{!}{\includegraphics{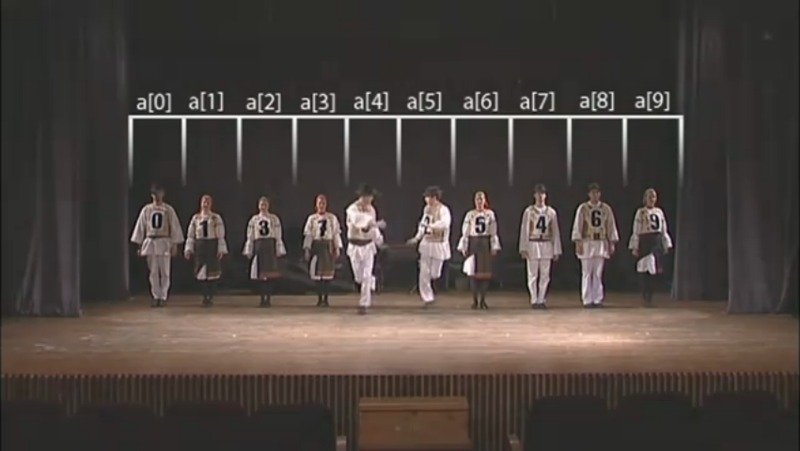}}
    \caption[]{Physical dance demonstration of a bubble sort algorithm~\cite{Zoltan2017}.}
    \label{fig:bubble-dance}
\end{figure}

\subsection{Exhibit 2: Data-Driven Performance}

Some storytelling takes place in person and is enacted by one or several performers. 
SketchStory by Lee et al.~\cite{Lee2013} (Figure~\ref{fig:sketch-story}) is close to traditional data-driven storytelling, and supports a person giving a presentation using an interactive whiteboard that responds to sketch input to generate interactive graphics.
However, SketchStory is curiously alone in this area---few visualization or data science tools exist to support in-person presentations. 
This was also the motivation for our GestureStory concept described later in this paper.

However, there are additional ways that people have used live performance to communicate data.
Figure~\ref{fig:bubble-dance} shows an image from a performance where dancers demonstrate how the bubble sort algorithm works.~\cite{Zoltan2017}
This algorithm pushes the largest element to the right and forms a ordered sequence of numbers.
For clarity, each dance is labeled in the this picture.
In reality, the dancers wear uniforms with numbers on them, where each dancer stands for a different element in the array.
The storyline is the movement of all the array elements.
The data is the ordered sequence.

A similar data-driven dance performance was created by the Dance.Draw project~\cite{LatulipeH08, Latulipe2011} (Figure~\ref{fig:dance.draw}), where the movements of dancers in a physical space was conveyed using visual representations. 
This mechanism could also be used as a vehicle for conveying data-driven stories.
Finally, the Data-Driven Dance project\footnote{\url{http://www.datadrivendance.org/}} (Figure~\ref{fig:arrastre}) has three performances designed to convey abstract data and blur the boundary between science and art: [arra]stre (data-driven dance for computer science theory), [data]storm (ocean storms, networks, and weather), and [pain]byte (biomedical engineering, chronic pain, and dance).

\textit{Informal classification:} 
In-person performances typically take place in an auditorium or studio, which supports a \textit{large audience}. 
Disregarding video recordings of the performance (which would be another form of media), this does require the audience to be physically \textit{co-located} with the performance, and to consume it in \textit{synchronously}, in real-time. 
This means that the performance is not stored; it is \textit{emphemeral}.
The components of the performance are human bodies in motion over the duration of the performance, and can also include text, sound, light, and visuals (typically projected in the workspace).

\begin{figure}[htb]
	\centering
    \resizebox{\columnwidth}{!}{\includegraphics{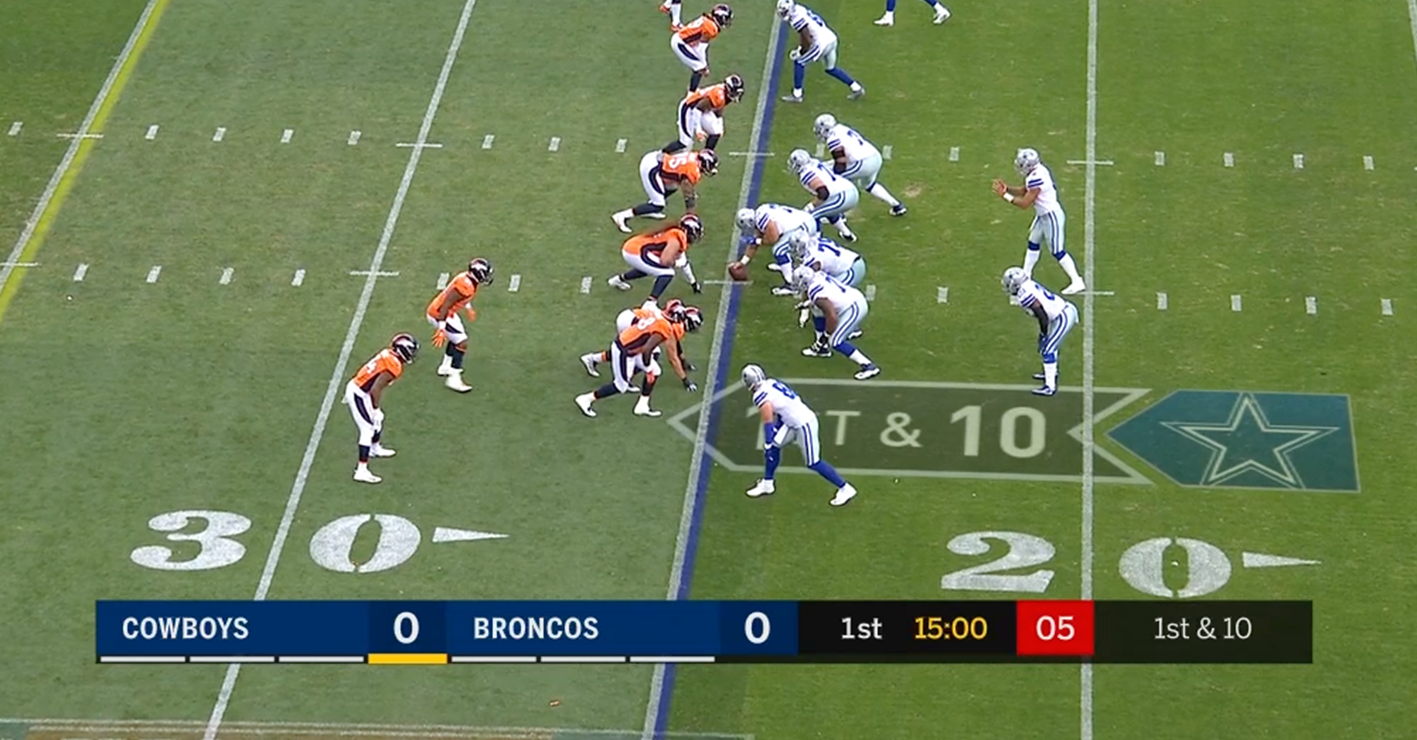}}
	\caption{Graphics displayed on the pitch for a football game.}
    \label{fig:sports}
\end{figure}

\begin{figure}[htb]
	\centering
    \resizebox{\columnwidth}{!}{\includegraphics{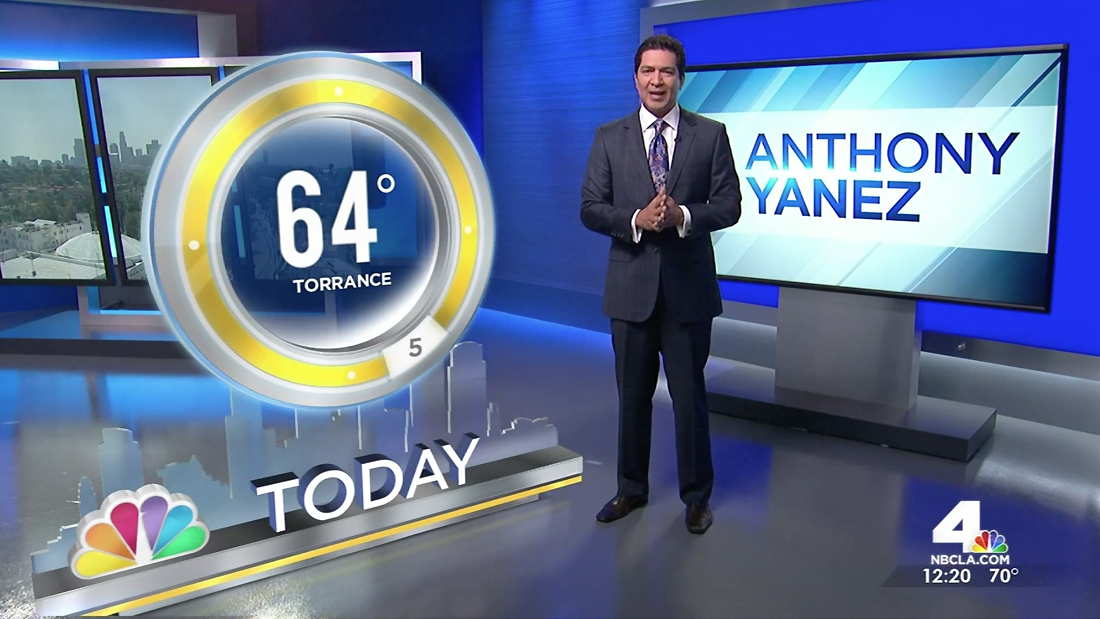}}
	\caption{Augmented reality forecast on NBC News.}
    \label{fig:nbc}
\end{figure}

\begin{figure}[htb]
	\centering
    \resizebox{\columnwidth}{!}{\includegraphics{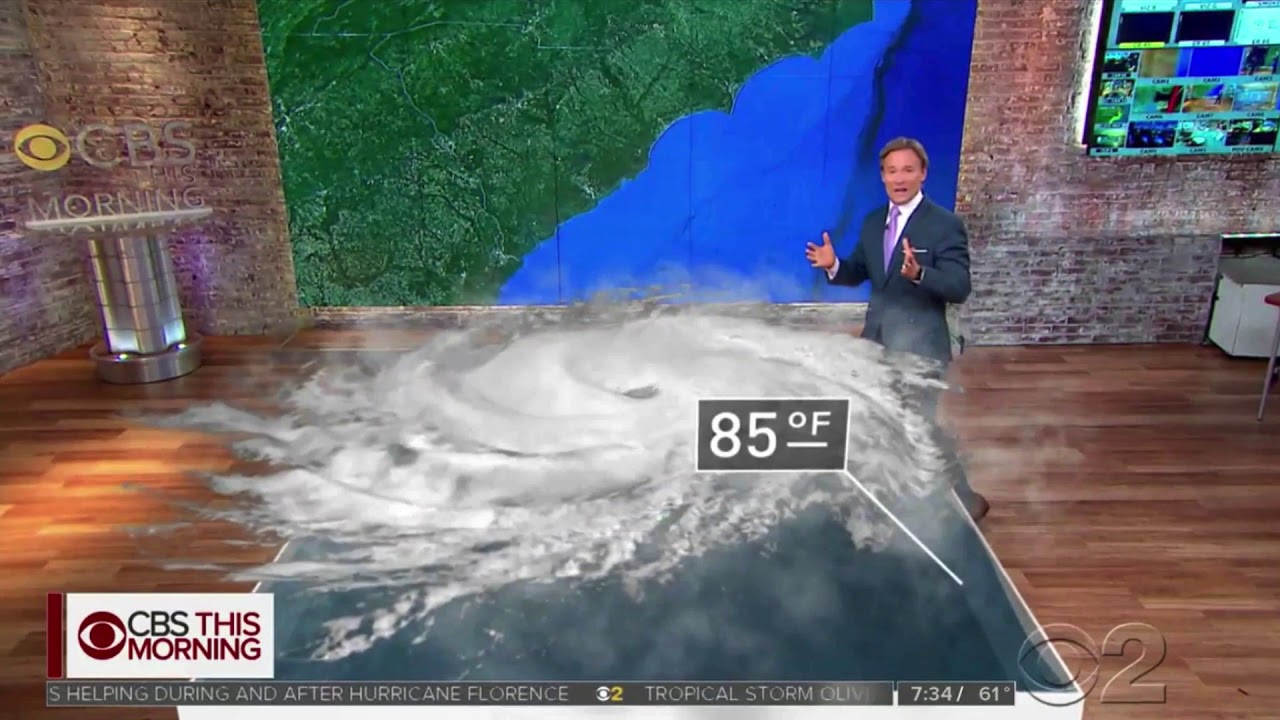}}
	\caption[]{Augmented reality forecast on CBS News.}
    \label{fig:cbs}
\end{figure}

\begin{figure}[htb]
	\centering
    \resizebox{\columnwidth}{!}{\includegraphics{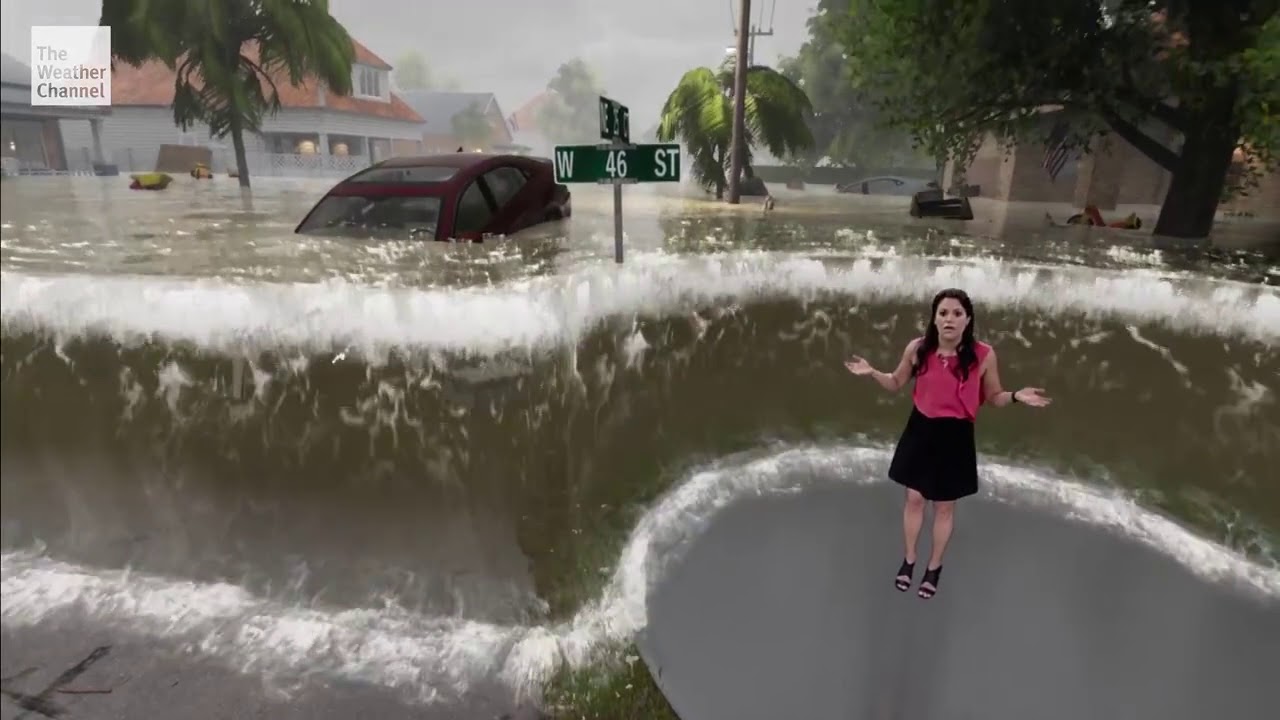}}
    \caption[]{Immersive augmented reality hurricane experience on the Weather Channel.}
    \label{fig:weather-channel}
\end{figure}

\subsection{Exhibit 3: Storytelling in Composite/Augmented Reality}

Augmented reality~\cite{Schmalsteig2016} is making significant inroads in society as technology becomes availability and new ideas to leverage the technology are introduced.
Aspects of technology has in fact already been used in a limited capacity in broadcasting for a long time, particularly in sports, where it can be most accurately named ``composite'' graphics.
Figure~\ref{fig:sports} shows a typical example of informative graphics inlaid on an American football field on a live broadcast.

The practice has also recently been employed in weather reporting. 
In fact, weather reporting has a long history of using image composition to combine the reporter's image with a virtual background through chromakeying.
Augmented reality allows for more advanced effects.
Figure~\ref{fig:nbc} shows a studio incorporating both virtual (the logo and temperature gauge) and real component (the weatherman) in the same shot. 
In Figure~\ref{fig:cbs}, this is taken to the next level by creating a 3D depiction of a hurricane in front of the reporter.
Finally, Figure~\ref{fig:weather-channel} shows the popular and much-discussed ``immersive hurricane'' footage from the Weather Channel, where the impact of Hurricane Florence on North Carolina in September 2018 was shown in a live broadcast.

\textit{Informal classification:} 
Clearly, augmented reality has much potential for storytelling, but its parameters depend on how it is used.
When experienced by a person using an AR headset, the audience is a single person, whereas the broadcasting examples discussed here are clearly intended for a broad audience.
Similarly, the interaction depends on whether it is experienced in person---in which case the person can navigate freely in the real world as well as interact with virtual objects---or viewed as composite graphics on a screen, where the interaction is limited to playback.

\subsection{Exhibit 4: Data-Driven Storytelling in Video Games}

While Segel and Heer~\cite{Segel2010} explicitly note that they chose not to include video games in their survey, games have long been instances where visualization is often integrated.~\cite{Bowman2013}
As it turns out, they have also been used for data-driven storytelling.
Figure~\ref{fig:civ3-replay} depicts an image from the replay session that is shown at the end of a completed game session, i.e., after one of the players (human or computer) has achieved one of the victory conditions: conquering all opponents, winning the space race, taking over most of the land, or scoring a diplomatic or cultural victory. 
The replay shows a history of how each civilization was founded, expanded across the world, and was eventually defeated.
While the interaction is limited, playback controls allows the user to go back and forward in the history.
Similar session playback tools---often called \textit{theater mode}---can be found in \textit{Halo 3} and \textit{Call of Duty: Black Ops}.

\textit{Informal classification:} 
The audience for most theater modes is \textit{individual} players who wants to study their own and other players' performance.
However, many theater modes typically also allow the player to cut and paste clips together, eventually producing a resulting \textit{video to share} with others.
The resulting video will have the same features as a data video (see above).
A playback session typically uses a map view, so the bandwidth requirement is lower than full-motion video, thus reducing the cognitive load.
One feature of most theater modes is that they make it easy to navigate in 2D or 3D in the scene, thus changing the viewpoint.
While the action itself cannot be changed (since it represents events that already happened), this interaction is powerful in that it can, for example, allow a player to put themselves in the shoes of another player to see what an encounter looked like from their viewpoint.

\begin{table*}[htb]
    \centering
    \includegraphics[width=\textwidth]{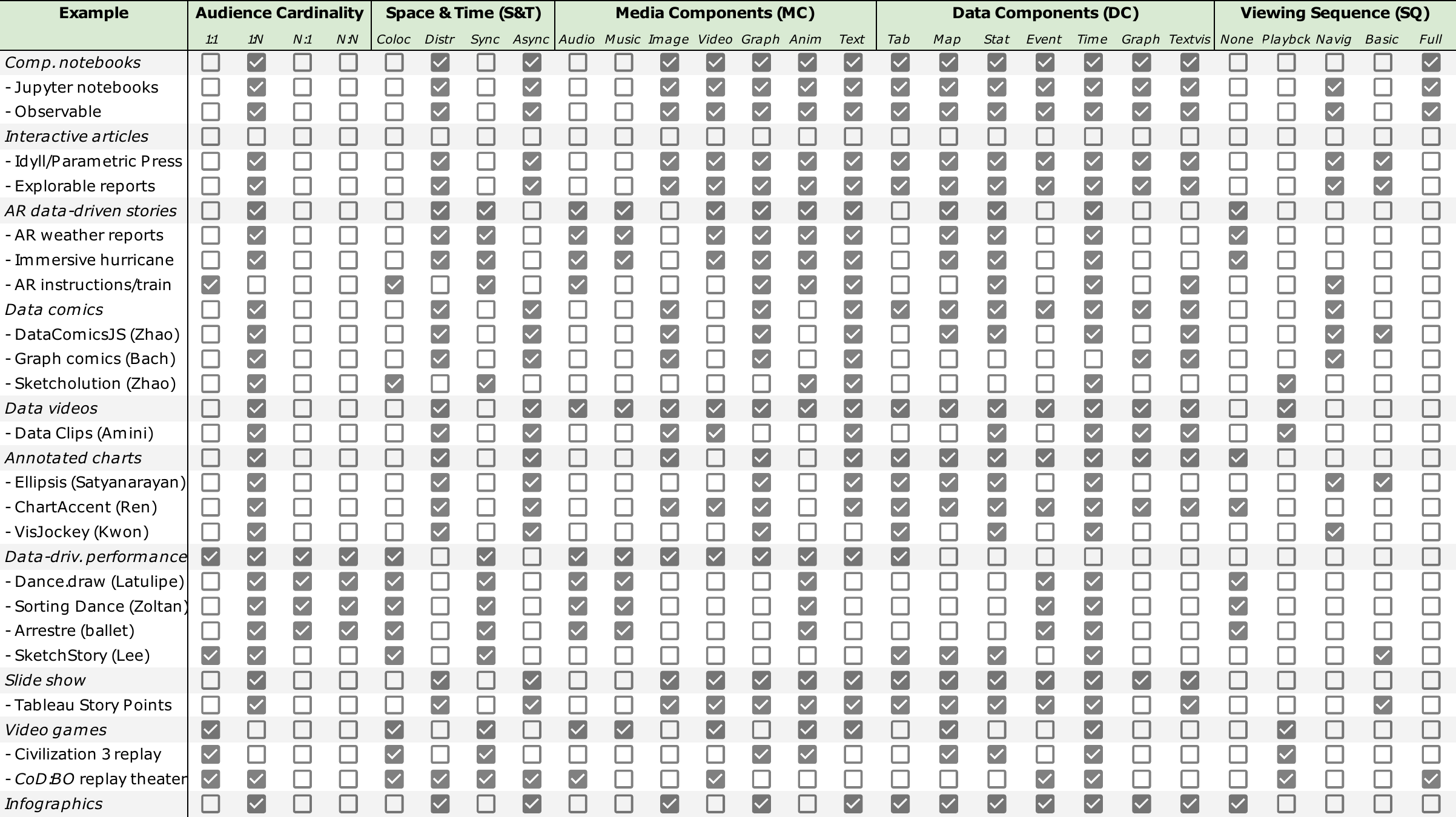}
	\caption{Classification of our representative sample of different media used for data-driven storytelling.
	Examples in italics are categories.
	}
    \label{tab:classification}
\end{table*}

\nocite{Bach2016, Bach2017}
\nocite{Zhao2019}
\nocite{Zhao:2015}
\nocite{Kwon2014, RenBLHC17, Kosara2013b, Kosara2013}
\nocite{SatyanarayanH14a}

\section{Taxonomy of Data-driven Storytelling Media}

In this paper, we propose to identify, study, and classify \textit{media} that have traditionally been used for storytelling.
The purpose of this activity is to expand existing genres of narrative visualization to encompass the entire scope of storytelling in society.
This, in turn, would generate a wealth of new research ideas for how to best use such media for data-driven storytelling.
Based on our survey of the existing evidence of a wide variety of media for data-driven storytelling, we found the following dimensions useful in classifying them:

\begin{itemize}
\item\textbf{Audience Cardinality (A):} Who is the intended recipient for the story?
\item\textbf{Space and Time (S/T):} What is the temporal and spatial delivery mechanism for the story?
\item\textbf{Media Components (VC):} What are the visual and sound building blocks employed?
\item\textbf{Data Components (DC):} How is the data conveyed to the viewer?
\item\textbf{Viewing Sequence (SQ)} How is the media viewed by the viewer?
\end{itemize}

The method we used to derive these dimensions included reviewing a large sample of examples of data-driven storytelling, and then narrowing our selection down a set of representative examples. 
These examples have been classified into the taxonomy and can be viewed in Table~\ref{tab:classification}.

\subsection{Audience Cardinality (A)}

In our notion of the audience of the data-driven story, we also include the storytellers: whether it is one or several people who are creating or viewing the narrative, respectively. 
A group performance, such as the bubble sort dance in the example above, would be an example of many storytellers (the dancers) conveying data to many recipients (the audience in the dance studio).

Audience has the below values:
\begin{itemize}
\item\textit{One-to-one} (1:1): one storyteller and one recipient, such as in a private conversation.
\item\textit{One-to-many} (1:N): one storyteller and many recipients, such as a speaker giving a talk to a group.
\item\textit{Many-to-one} (N:1): many storytellers and one recipient, such as an entire staff briefing a commander.
\item\textit{Many-to-many} (N:N): many storytellers and many recipients, such as a dance troupe giving a performance to an audience.
\end{itemize}

\subsection{Space and Time (S/T)}

We borrow the notion of space and time from CSCW,~\cite{Baecker1993} where the space-time matrix has long been used to characterize forms of groupware based on the spatial and temporal relations of the human users. 
It is a useful property because both space and time have a significant impact on the delivery and storage mechanism for the data-driven story.

Space and time has two values, one for each dimension:
\begin{itemize}
\item\textit{Space}: relative physical locations of storyteller and recipient.
	\begin{itemize}
    \item\textit{Co-located} (coloc): the storyteller and the recipient are in the same physical space.
    \item\textit{Distributed} (distr): the storyteller and the recipient are \textbf{not} in the same physical space.
    \end{itemize}
\item\textit{Time}: temporal locations of storyteller and recipient.
	\begin{itemize}
    \item\textit{Synchronous} (sync): the storyteller is delivering the story to the recipient in real time.
    \item\textit{Distributed} (distr): the storyteller is delivering the story in a form that the will be consumed by the recipient at a later time.
    \end{itemize}
\end{itemize}

\subsection{Media Components (MC)}

The nature of the media captures the composition of the media being used for data-driven storytelling.
Since we are typically talking about composite media types, the Media Components is a set variable that can include one or several of the below:

\begin{itemize}
\item\textit{Audio} (aud): audio, such as speech recordings, ambient noise, or sampled sound effects.
\item\textit{Music} (mus): ordered sound forming a musical piece.
\item\textit{Photographs} (pho): pixmap images.
\item\textit{Live video} (vid): animated pixmap images.
\item\textit{Static graphics} (gra): non-dynamic vector graphics.
\item\textit{Animated graphics} (ani): dynamic vector graphics.
\item\textit{Text} (txt): textual representations.
\end{itemize}

\subsection{Data Components (DC)}

The core purpose of a data-driven storytelling artifact is to convey data from the storyteller to the viewer.
The form that this takes is captured in the Data Components (DC) dimension. 
It is a set variable that will take or several of the below values:

\begin{itemize}
\item\textit{Table} (tab): data table representation;
\item\textit{Map} (map): spatial maps;
\item\textit{Statistical graphics} (stat): classic statistical graphics such as barcharts, scatterplots, and linecharts;
\item\textit{Discrete event visualization} (event): event timelines;
\item\textit{Continuous time visualization} (time): time-series data;
\item\textit{Graph visualization} (graph): network visualizations, such as node-link or adjacency matrix representations; and
\item\textit{Text visualization} (txtvis): textual visualizations, such as wordclouds or visual concordances.
\end{itemize}

\subsection{Viewing Sequence (SQ)}

The level of interactivity associated with a storytelling artifact governs its level of engagement, cognitive load, and adaptiveness. 
This variable takes one of the following values:

\begin{itemize}
\item\textit{No interaction} (noint): the artifact cannot be interacted with.
It will play out from start to finish.
\item\textit{Playback control} (playback): the viewer can stop and rewind or at least restart the narrative.
\item\textit{Navigation} (nav): the artifact allows the user to zoom and pan around in the representation.
\item\textit{Basic control} (basic): the artifact allows the user to click and focus on part of the representation to highlight or trigger other effects.
\item\textit{Full control} (control): the artifact yields full filtering, linking, and transformation control to the user.
\end{itemize}




\section{Applications}

Given our taxonomy, we here explore a few designs for data-driven storytelling in detail.

\subsection{Sequential Art for Data: Data Comics}

Data comics is an approach to show how \textit{sequential art}~\cite{McCloud1994}---also known as comics---can be used as a novel method for storytelling.
Several researchers have studied this phenomenon, including Zhao et al.,~\cite{Zhao:2015, Zhao2015, Zhao2019} Bach et al.,~\cite{Bach2016, Bach2017} and Kim et al.~\cite{Kim2019}
We here discuss a practical system for building data comics.

This data comics authoring system allows the users to build narratives using comic layouts of panels that contain both images, text, figures, and live visualizations.
Comic features, such as motion lines, captions, thought and speech bubbles, build on the universal language of comics.
Our corresponding implementation provides an authoring system for creating a data comic by clipping images, data, and visualizations using a web browser and combining them into the comic layout.

\begin{figure}[htb]
  \centering
  \resizebox{\columnwidth}{!}{\includegraphics{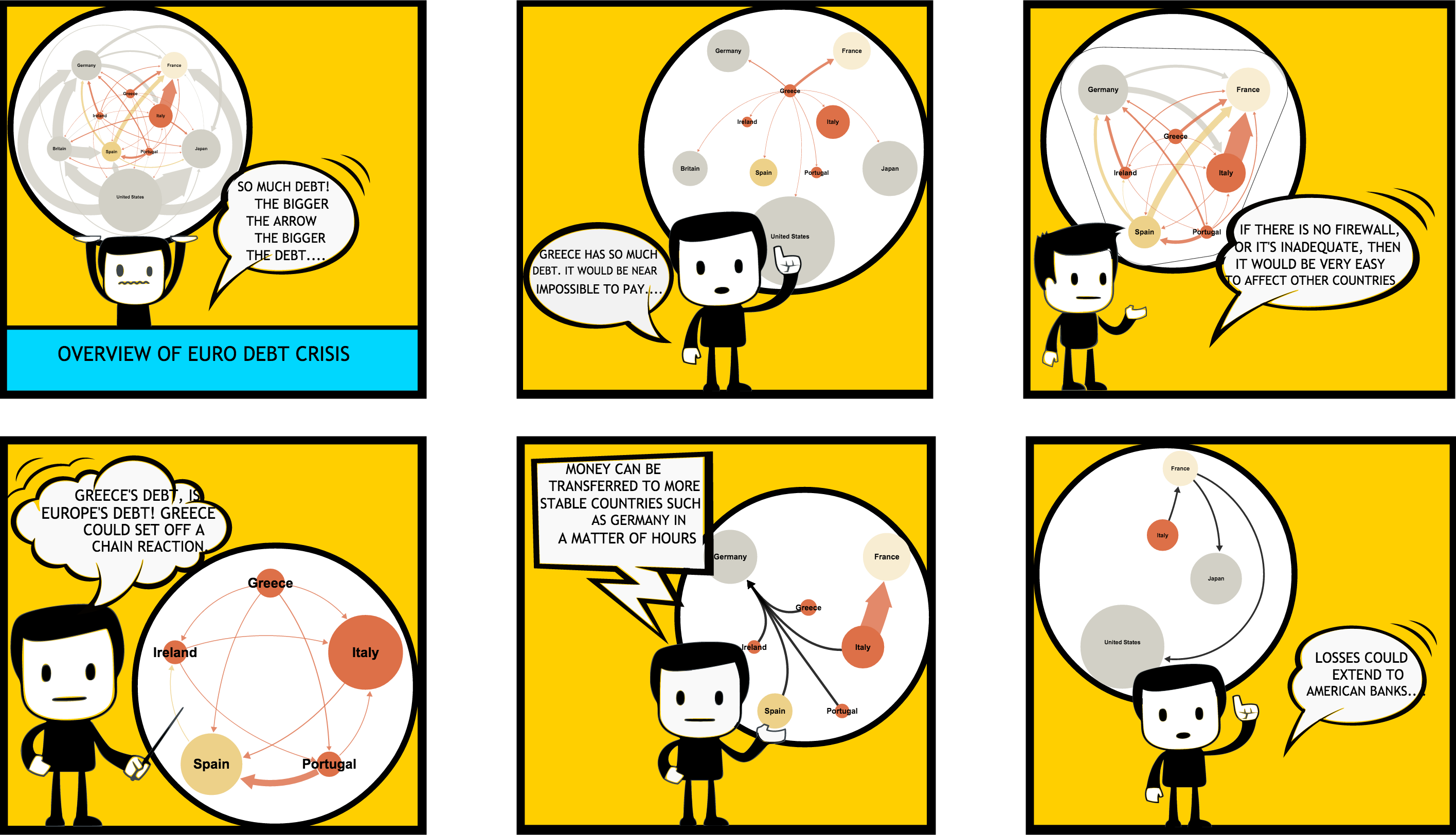}}
  \caption{Data Comic on the European debt crisis.}
  \label{fig:eurodebt}
\end{figure}

Figure~\ref{fig:eurodebt} is an example of a data comic produced using our prototype implementation by clipping sources and images from the web. 
More specifically, the designer starts by collecting the data visualizations and other material.
The designer then crops subgraphs and splits them across panels to allow the reader to focus on one part at a time (rather than seeing the entire thing at once).

In terms of our taxonomy, Data Comics is designed for both entertainment, information, and education. 
The author cardinality (A) is one-to-many, since that a comics is typically designed to be viewed on a computer or personal device, and thus it has a distributed and asynchronous space and time (S/T) value.
Comics are generally not interactive, but they do support navigation. 
The media components include static graphics, and text, whereas the data components include virtually any static visualization element.
Comics are permanent in that they can be stored and transmitted easily.

\subsection{Live-Streaming Data Analysis: DataTV}

\begin{figure}[htb]
  \centering
  \frame{\resizebox{\columnwidth}{!}{\includegraphics{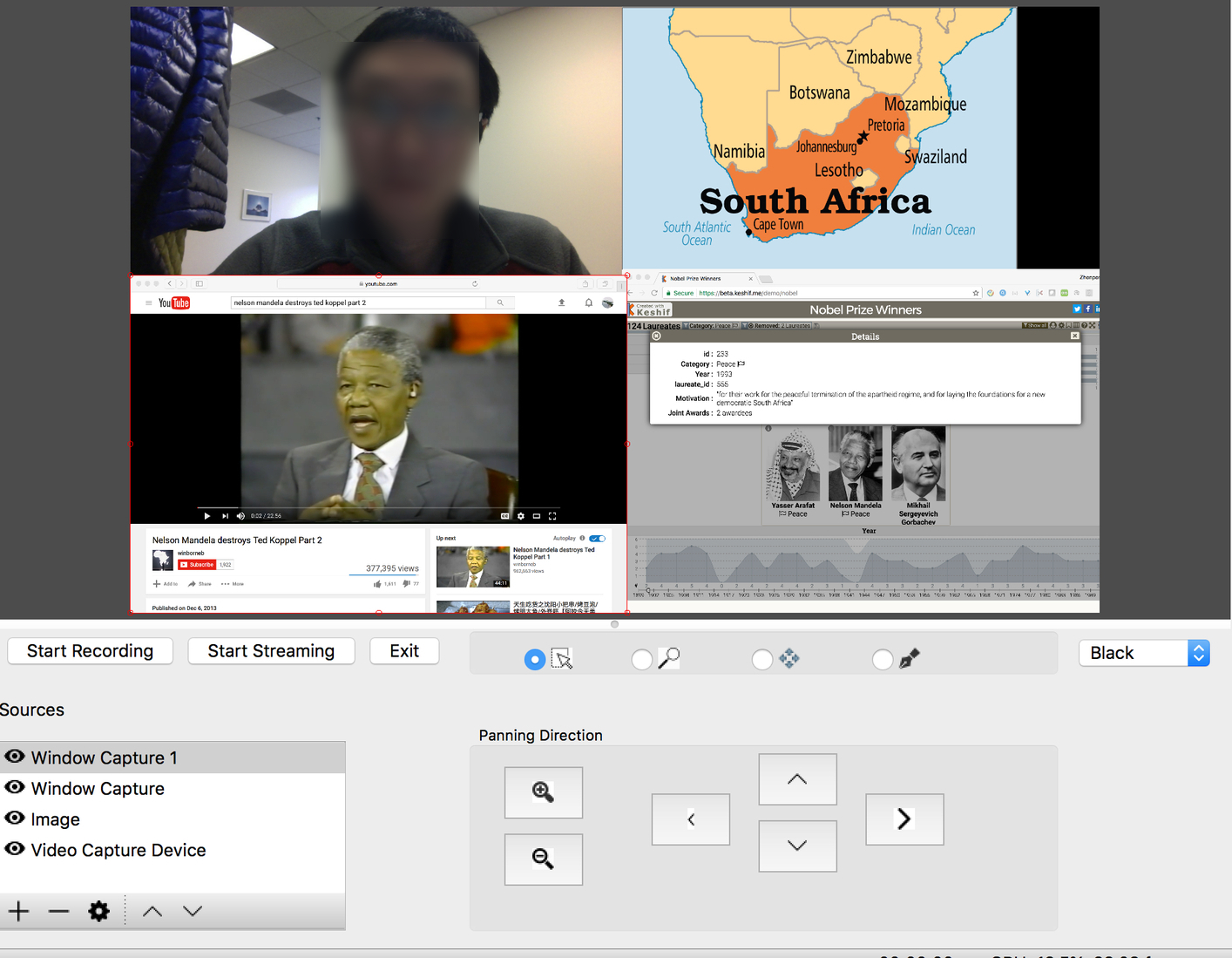}}}
  \caption{Webcam and Keshif visualization being recorded for the Nobel Prize Winner data video.
  A data video can be made with user as the presenter.}
  \label{fig:keshif_nobel_interactive}
\end{figure}

DataTV is a prototype system for authoring live-streaming data videos using a single, integrated desktop interface.
The approach is based on the notion that live-streamed video has not yet been explored for data-driven storytelling.
The prototype (Figure~\ref{fig:keshif_nobel_interactive}) supports three separate modes for (1) production, (2) recording, and (3) editing, in a highly streamlined and optimized workflow that allows a single content creator to control the entire process, even during live streaming.
The tool incorporates multimedia sources such as live webcams, live audio recordings, web browsers, image viewers, and full-motion video.
In particular, it supports live recording of any selected window on the user's desktop, such as those containing an interactive visualization, such as a web browser or dedicated application window.
Furthermore, the tool incorporates advanced video functionality, such as chroma-keying (making parts of a stream transparent, such as for blue or green screens), picture-in-picture, hand-drawn annotations (for highlighting important parts of a stream), viewport control (zooming and panning), and advanced source composition operations (transitions, stretching, and fitting).

In terms of our taxonomy, Data TV is designed for both entertainment, information, and education. 
The author cardinality (A) is one-to-many, since that a live streaming is typically designed to be viewed on a computer or personal device, and thus it has a distributed and synchronous space and time (S/T) value.
Live streaming generally does not support interaction (unless the streamer provides a chat channel for viewers to give feedback), so the viewing sequence is playback only.
The media components include audio, live video, static graphics, and text, whereas the data components include virtually any static visualization element.
Live streaming is generated once, and the interaction cannot be replicated.

\subsection{Supporting Oral Tradition: GestureStory}

Inspired by our taxonomy and its gaps, we propose a third application for data-driven storytelling: body language and gestures.
Body language has long been a part of storytelling, but it is only recently where modern technology has enabled people to use their gestures and entire bodies to control digital devices.
What if we could use data-driven storytelling to support a speaker talking to an audience, simply by using the speaker's body language and gestures as input to control the audio-visual material (slides and data visualizations) used to support the presentation?
Inspired by Lee et al.'s work on SketchStory,~\cite{Lee2013} where pen input drives the narrative, we call this idea GestureStory.

A GestureStory tool would enable the presenter or even the audience to control the visual components of a presentation consisting of data visualizations using their gestures and voices.
The application would presumably use a depth camera and microphone to recognize body movements and spoken words.
These can be used to control interactive operations such as panning, zooming, moving, annotating, and etc.
The visual components can be made to follow the gesture controls while textual annotations can be added by transcribing the presenter's speech. 

In terms of our taxonomy, a gesture story tool is designed for entertainment, information, and education.
The author cardinality is one-to-many, since a presentation is typically designed to be viewed in the same location as the presenter while being presented, and thus it has a distributed and synchronous space and time (S/T) value.
The benefit of a live presentation is that the audience can interact with the presenter by asking questions, so this media supports at least basic interaction control.
The media components include audio, static graphics, animation, and text, whereas the data components include table most of the static visualization element.
Gesture stories are generated with the author's improvisations, and thus can not be replicated easily.
We classify it as part of our ``data-driven performance'' category.

\section{Implications for Design}

We have successfully labeled some thirty representative examples using our taxonomy (see Table~\ref{tab:classification}).
The examples cover movies, documentaries, web articles with data visualization, infographics, comics, social media, visualization tools, games, dance, and sketching tools, which are a large part of the major categories for storytelling.
This comprehensive classification provides good evidence that our taxonomy is sufficient and complete. 

Our taxonomy is an extension that builds on the foundation that Segel and Heer~\cite{Segel2010} laid in 2010.
While this foundation has proven instrumental in the guiding the development of data-driven storytelling, their model is limited in scope and conflates the delivery mechanism with the media used for the message.
We believe our taxonomy provides a more comprehensive view of media for data-driven storytelling while still building on their foundational work.
Using our taxonomy, designers will be able to widen the horizon of data-driven storytelling.
By providing a taxonomy with detailed dimensions, we explored the possible values for each dimension. By expanding the list of dimensions and dimensions values, we can also keep tracking of the emerging media types. 
In particular, the terms and dimensions in our taxonomy provides a standardized vocabulary to use when discussing data-driven storytelling.
This enables researchers and practitioners alike to classify their own work so that existing and new media can be systematically organized with a common ground.

However, the true value of a taxonomy such as ours is in generating new ideas by identifying gaps in the literature.
By grouping and labeling the dimensions of existing media, our taxonomy can help researchers identify new areas to explore in the future.
For example, this new design space can be generated by exploring previously untested combinations of dimensions.
We have done so in the previous section: the Data Comics authoring system as well as the Data TV and GestureStory applications are all based on our taxonomy.

A common theme for many of the novel storytelling methods such as Distill, Idyll, and even some data comics is \textit{interactivity}. 
This clearly goes in hand with the old adage attributed to Confucius that \textit{''I hear and I forget/I see and I remember/I do and I understand.''}
In other words, learning is best scaffolded by interactivity that allows the learner to change parameters and study its effects.
An concrete example is Omar Shehata's Parametric Press article ``Unraveling the JPEG''\footnote{\url{https://parametric.press/issue-01/unraveling-the-jpeg/}}, which provides many interactive tools for exploring the surprising depths of the JPEG image format.
It is clear that the future of data-driven storytelling will include interactive media.

A particular such interactive medium is video games. 
We have already surveyed some examples of data-driven storytelling in games (or created through the use of games in what is known as \textit{Machinima}), and we think that there is significant potential for this in the future. 
In particular, games for education and/or games in augmented reality may be particularly fertile grounds for adoption data-driven storytelling.

Let us close on Marshall McLuhan's famous note that ``\textit{the medium is the message}.''~\cite{McLuhan1964}
It is clear that the medium used to convey a story will also affect the message and content of that story.
This is not only true because the medium we choose affects the audience that will be able to consume it, but also that the unique aspects of the medium gives specific capabilities to the storyteller.
We look forward to seeing how the data-driven storytelling community will explore such novel media in the future.

\section{Conclusion and Future Work}

We have proposed a taxonomy of media for data-driven storytelling for the purpose of widening the discourse on which storytelling techniques can be used for telling stories about data.
Our work began with a survey of the wide range of evidence of such novel data-driven stories, resulting in us identifying several representative dimensions to categorize these media types.
We then defined our taxonomy and used it to classify a large collection of novel data stories.
Based on these ideas, we propose several novel applications for data-driven storytelling, and then close the paper with some take-aways for designers looking to expand their repertoire in this domain.

In the future, we anticipate continuing our investigations into novel methods for data-driven storytelling.
We believe that this taxonomy will be helpful in guiding our efforts in this endeavor.
	
\section*{Acknowledgments}

This work was partially supported by the U.S.\ National Science Foundation award IIS-1539534.
Any opinions, findings, and conclusions or recommendations expressed in this material are those of the authors and do not necessarily reflect the views of the funding agency.

\bibliographystyle{plainnat}
\bibliography{storytelling-media}

\end{document}